\documentclass{aastex62}
\usepackage{amsmath,amssymb,CJK,soul}

\received{--}
\revised{--}
\accepted{--}
\submitjournal{AJ}

\shorttitle{A Twilight Search for Atiras, Vatiras and Co-orbital Asteroids}
\shortauthors{Ye et al.}

\begin{document}
\begin{CJK*}{UTF8}{gbsn}

\title{A Twilight Search for Atiras, Vatiras and Co-orbital Asteroids: Preliminary Results}

\correspondingauthor{Quanzhi Ye}
\email{qye@umd.edu}

\author[0000-0002-4838-7676]{Quanzhi Ye (叶泉志)}
\affiliation{Division of Physics, Mathematics and Astronomy, California Institute of Technology, Pasadena, CA 91125}
\affiliation{IPAC, California Institute of Technology, 1200 E. California Blvd, Pasadena, CA 91125}
\affiliation{Department of Astronomy, University of Maryland, College Park, MD 20742}

\author{Frank J. Masci}
\affiliation{IPAC, California Institute of Technology, 1200 E. California Blvd, Pasadena, CA 91125}

\author{Wing-Huen~Ip \begin{CJK*}{UTF8}{bkai}(葉永烜)\end{CJK*}}
\affiliation{Institute of Astronomy, National Central University, 32001, Taiwan}
\affiliation{Space Science Institute, Macau University of Science and Technology, Macau}

\author{Thomas A. Prince}
\affiliation{Division of Physics, Mathematics and Astronomy, California Institute of Technology, Pasadena, CA 91125}

\author{George Helou}
\affiliation{IPAC, California Institute of Technology, 1200 E. California Blvd, Pasadena, CA 91125}

\author{Davide Farnocchia}
\affiliation{Jet Propulsion Laboratory, California Institute of Technology, 4800 Oak Grove Drive, Pasadena, CA 91109}

\author[0000-0001-8018-5348]{Eric C. Bellm}
\affiliation{DIRAC Institute, Department of Astronomy, University of Washington, 3910 15th Avenue NE, Seattle, WA 98195}

\author{Richard Dekany}
\affiliation{Caltech Optical Observatories, California Institute of Technology, Pasadena, CA 91125}

\author[0000-0002-3168-0139]{Matthew J. Graham}
\affiliation{Division of Physics, Mathematics, and Astronomy, California Institute of Technology, Pasadena, CA 91125}

\author[0000-0001-5390-8563]{Shrinivas R. Kulkarni}
\affiliation{Division of Physics, Mathematics, and Astronomy, California Institute of Technology, Pasadena, CA 91125}

\author[0000-0002-6540-1484]{Thomas Kupfer}
\affiliation{Kavli Institute for Theoretical Physics, University of California, Santa Barbara, CA 93106}

\author[0000-0003-2242-0244]{Ashish Mahabal}
\affiliation{Division of Physics, Mathematics, and Astronomy, California Institute of Technology, Pasadena, CA 91125}
\affiliation{Center for Data Driven Discovery, California Institute of Technology, Pasadena, CA 91125}

\author[0000-0001-8771-7554]{Chow-Choong Ngeow}
\affiliation{Institute of Astronomy, National Central University, 32001 Taiwan}

\author{Daniel J. Reiley}
\affiliation{Caltech Optical Observatories, California Institute of Technology, Pasadena, CA 91125} 

\author[0000-0001-6753-1488]{Maayane T. Soumagnac}
\affiliation{Department of Particle Physics and Astrophysics, Weizmann Institute of Science, Rehovot 76100, Israel}
\affiliation{Lawrence Berkeley National Laboratory, 1 Cyclotron Road, Berkeley, CA 94720}



\begin{abstract}

Near-Earth Objects (NEOs) that orbit the Sun on or within Earth's orbit are tricky to detect for Earth-based observers due to their proximity to the Sun in the sky. These small bodies hold clues to the dynamical history of the inner solar system as well as the physical evolution of planetesimals in extreme environments. Populations in this region include the Atira and Vatira asteroids, as well as Venus and Earth co-orbital asteroids. Here we present a twilight search for these small bodies, conducted using the 1.2-m Oschin Schmidt and the Zwicky Transient Facility (ZTF) camera at Palomar Observatory. The ZTF twilight survey operates at solar elongations down to $35^\circ$ with limiting magnitude of $r=19.5$. During a total of 40 evening sessions and 62 morning sessions conducted between 2018 November 15 and 2019 June 23, we detected 6 Atiras, including 2 new discoveries 2019 AQ$_3$ and 2019 LF$_6$, but no Vatiras or Earth/Venus co-orbital asteroids. NEO population models show that these new discoveries are likely only the tip of the iceberg, with the bulk of the population yet to be found. The population models also suggest that we have only detected 5--$7\%$ of the $H<20$ Atira population over the 7-month survey. Co-orbital asteroids are smaller in diameters and require deeper surveys. A systematic and efficient survey of the near-Sun region will require deeper searches and/or facilities that can operate at small solar elongations.

\end{abstract}



\section{Introduction}

After three decades of regular asteroid surveying, our knowledge of the near-Earth object (NEO) population has vastly improved: more than $95\%$ of kilometer-class NEOs have been cataloged \citep{Jedicke2015}. Next-generation sky surveys, such as the Large Synoptic Survey Telescope \citep{Schwamb2018} and NEOCam \citep{Mainzer2018}, will survey the visible sky $\sim10\times$ deeper than current assets, and will likely push the completeness into the sub-kilometer \citep[e.g.][and references therein]{Chesley2017}.

However, because of selection effects, there are a few asteroid groups that remain poorly characterized, the most notable are the asteroids that orbit the Sun on or within the Earth's orbit. These asteroids spend most of their time near the Sun as seen from Earth, making them difficult to detect. Most objects that fall into this class are known as Atiras or Interior-Earth Objects \citep[originally known as the Apoheles; c.f.][]{Tholen1998, Binzel2015b}, whose orbits are entirely confined by the Earth's orbit\footnote{Technically speaking, Atiras are defined as the objects with an aphelion $Q$ smaller than Earth's perihelion distance, i.e. $Q<0.983$~au. However, some objects have orbits with $Q>0.983$~au but are still entirely confined within Earth's orbit due to the eccentricity of the latter. These asteroids are known as ``pseudo-Atiras'' and are technically not Atiras.}. Others are Earth-co-orbital objects \citep[c.f.][]{Brasser2004}, which have an orbit similar to that of the Earth. The definition of the different orbital types investigated in this paper is summarized in Table~\ref{tbl:def}.

\begin{table*}
\begin{center}
\caption{Definition of different NEO orbital types investigated in this paper. Definition of orbital elements are: $a$ -- semimajor axis, $e$ -- eccentricity, $i$ -- inclination, $q$ -- perihelion distance, $Q$ -- aphelion distance.\label{tbl:def}}
\begin{tabular}{ll}
\hline
Type & Definition \\
\hline
NEOs & $q<1.3$~au and $a<4.2$~au \\
Atiras & NEOs with $Q<0.983$~au \\
Pseudo-Atiras & NEOs with $0.983$~au$<Q<1.017$~au, but with orbits that are still confined by Earth's orbit \\
Vatiras & NEOs with $Q<0.718$~au \\
Earth co-orbitals & $0.998$~au$<a<1.002$~au, $e<0.3$, $i<30^\circ$ \\
Venus co-orbitals & $0.721$~au$<a<0.725$~au, $e<0.3$, $i<30^\circ$ \\
\hline
\end{tabular}
\end{center}
\end{table*}

Virtually all objects in the Atira region originated from the main asteroid belt and only reached their current orbits due to planetary perturbations and non-gravitational (e.g. Yarkovsky) effects \citep[c.f.][]{Granvik2018a}. These effects can continue to lower the perihelion distances of Atiras, making them more susceptible to thermally-driven disintegration due to intense heating in the solar environment \citep{Granvik2016a, Ye2019a}.

The Atira region is generally dynamically chaotic, but low eccentricity Atiras and co-orbital asteroids are more stable \citep{Ribeiro2016}. A very small but important sub-population is the asteroids residing near the Sun-Earth/Venus Lagrange L$_4$ and L$_5$ points, known as the Earth/Venus Trojans (ETs/VTs). Dynamical simulation shows that these objects can survive over timescales comparable to the age of the solar system \citep{Tabachnik2000}, implying that a small, ancient population of asteroids may exist in these regions. There is only one known ET \citep[2010 TK$_7$;][]{Connors2011}, and no VT has been detected yet, but the recently-detected zodiacal dust ring on the orbit of Venus seems to support the existence of Venus co-orbital asteroids \citep{Pokorny2019c}. The population of co-orbital asteroids, if exists, is very small: \citet{Morais2002, Morais2006} find that the population of Venus and Earth co-orbitals has only 0.1 1-km-class object and 4 100-m-class objects for Venus, and 0.6 1-km-class and 16 100-m-class objects for Earth, respectively. The true numbers are uncertain due to the lack of observational constraints and the poorly-constrained non-gravitational effects \citep{FuenteMarcos2014, Malhotra2019}.

\begin{figure*}
\includegraphics[width=\textwidth]{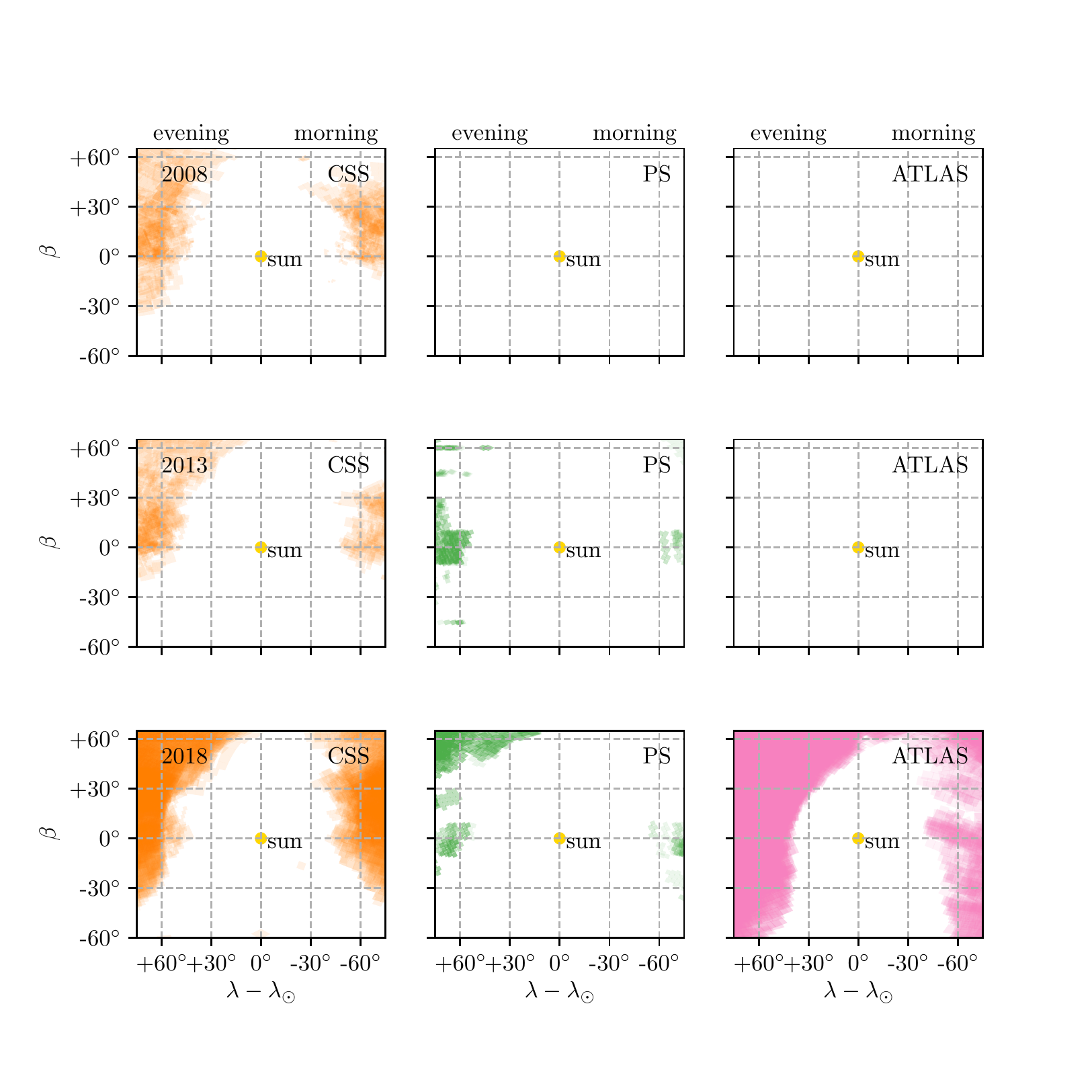}
\caption{Sky coverage of major NEO surveys in 2008, 2013 and 2018 -- the Catalina Sky Survey (CSS) system, PANSTARRS, and Asteroid Terrestrial-impact Last Alert System (ATLAS). An animation of this figure is available on the journal and shows the sky coverage of these three surveys from 2006 to 2018. Deeper colors represent more frequent visits.}
\label{fig:covmap-others}
\end{figure*}

For Earth-based observers, Atiras and co-orbital asteroids are generally only observable in the brief windows during evening and morning twilights. The earliest documented effort of asteroid searches during the twilight hours is the Trojan Vulcan (Mercury Trojans) survey, conducted by R. Trumpler and others, in the 1900s \citep{Trumpler1923a}. More recently, two short-term twilight surveys were conducted by \citet{Whiteley1998} and \citet{Ye2014}, who used the University of Hawaii's 2.24-m telescope (UH88) and the 3.6-m Canada-France-Hawaii Telescope (CFHT) to search for NEOs and sungrazing comets. The UH88 survey found a few NEOs, including the first Atira 1998 DK$_{36}$. The CFHT survey did not find any comets. Other than these short-term surveys, typical NEO surveys tend to stay out of the low elongation region, although one of the surveys \citep[Asteroid Terrestrial-impact Last Alert System, ATLAS;][]{Heinze2017} has recently expanded its coverage of the sky down to $45^\circ$ from the Sun (Figure~\ref{fig:covmap-others}).

Here we present a survey for Atiras and co-orbital asteroids, conducted using the recently commissioned Zwicky Transient Facility (ZTF) Camera, operated on the 1.2-m Oschin Schmidt telescope at Palomar Observatory. With 30-second exposures, ZTF can observe an area of 3760 deg$^2$ per hour down to $r=20.7$, making it an ideal instrument to search over a large area in the brief twilight window.

\section{The ZTF Twilight Survey} \label{sect:ztf-ts}

The ZTF survey is an optical, wide-field survey utilizing the 1.2-m Oschin Schmidt at Palomar Observatory. It has a 576 mega-pixel camera that was built specifically for the survey. The camera has a 55 deg$^2$ field-of-view and a fill-factor of $87\%$, corresponding to $\sim$47 deg$^2$ field-of-view. The survey is comprised of two baseline surveys and a number of mini-surveys, with the twilight survey being one of the latter. The hardware, survey strategy and science goals are described in detail in \citet{Bellm2019} and \citet{Graham2019}.

The ZTF twilight survey (ZTF-TS hereafter) uses both evening and morning $12^\circ$--$18^\circ$ twilight time to search for Atiras and co-orbital asteroids. The survey initially ran every night from 2018 November 17 to 2019 February 12. The cadence was reduced to one run every three nights from 2019 March 30 to 2019 June 23. Before each night's observations, the ZTF scheduler \citep{Bellm2019a} selects 10 ZTF survey fields that have smallest solar elongations with airmass $<2.92$ (i.e. a minimum elevation of $20^\circ$) throughout the twilight time and assigns these fields to the observation queue. The minimum solar elongation that can be observed under this setting is the sum of the maximum angular distance from the Sun (below the horizon) to the horizon and the minimum elevation angle, minus one half of the field width, i.e. $18^\circ+20^\circ-7^\circ/2=34.5^\circ$. Each field is imaged four times in ZTF $r$-band separated by $\sim5$~minutes in order to facilitate moving object detection, resulting a total survey time of $\sim20$--$25$~minutes every session.

Data from ZTF-TS are processed by the ZTF Science Data System \citep[c.f.][]{Masci2019}. The ZTF Science Data System employs two separate algorithms to detect moving objects: {\tt ZSTREAK} looks for fast-moving objects with apparent motions larger than $10^\circ$/day \citep{Duev2019, Ye2019}; ZTF Moving Object Discovery Engine ({\tt ZMODE}) looks for slower objects $<10^\circ$/day that show up as point-like objects in the images \citep{Masci2019}. Moving objects identified in the process are then compared against the catalog of known asteroids, using a local version of Bill Gray's {\tt astcheck}\footnote{\url{https://www.projectpluto.com/pluto/devel/astcheck.htm}}. Known asteroids in the survey data are typically bright main-belt asteroids near solar conjunction. We note that {\tt astcheck} examines known objects in the entire field-of-view, including those in the chip gaps that account for $13\%$ of the total imaging area. Candidates that do not match any known asteroids are then presented to a human operator for inspection. Candidates that are judged to be real asteroids are submitted to the Minor Planet Center (MPC)\footnote{\url{https://www.minorplanetcenter.net/}}. The MPC will then perform its own check against known asteroids, and the likelihood of the object being a NEO \citep[e.g.][]{Keys2019}. Unknown asteroids that are likely to be NEOs will be posted on the Near-Earth Object Confirmation Page (NEOCP)\footnote{\url{https://www.minorplanetcenter.net/iau/NEO/toconfirm_tabular.html}} to encourage follow-up observations.

\section{Results}
\label{sec:results}

\begin{figure*}
\includegraphics[width=0.5\textwidth]{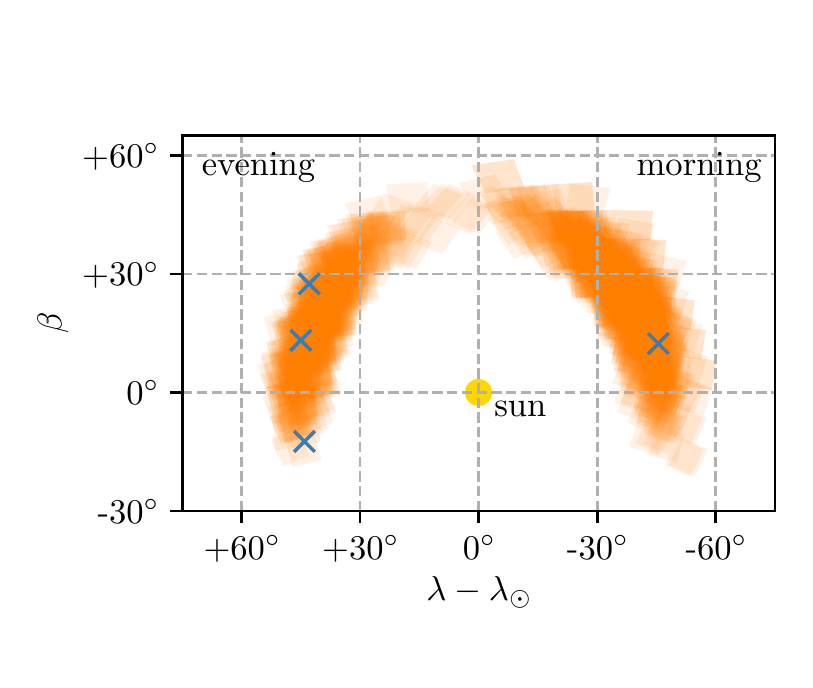}
\caption{Coverage map of the ZTF twilight survey plotted under Sun-centered ecliptic frame. Deeper color corresponds to more frequent visits. The discovery locations of 2019 AQ$_3$, 2019 LL$_5$, 2019 LF$_6$ and ZTF031C are also shown.}
\label{fig:covmap}
\end{figure*}

ZTF-TS was conducted in a total of 82 nights from 2018 November 17 to 2019 June 23, including 40 evening sessions and 62 morning sessions (some nights have only evening or morning sessions). The observed fields have solar elongation of $35^\circ$--$60^\circ$, as illustrated in Figure~\ref{fig:covmap}. We then use the function defined in \citet{Denneau2013} to derive the limiting magnitude of the survey based on the detection statistics of known asteroids. The detection efficiency is defined as

\begin{equation}
\label{eq:func}
\epsilon=\frac{\epsilon_0}{1+\exp{[(V-L)/W]}}
\end{equation}

\noindent where $\epsilon_0$ is the maximum efficiency among the bins, $V$ and $L$ are the apparent and limiting $V$-band magnitudes, respectively, and $W$ represents the characteristic ``width'' of the transition from maximum to zero efficiency. The $V$ magnitudes are quoted from the output of {\tt astcheck} (which gives asteroid brightness in $V$) and does not involve magnitude transformation. We note that {\tt astcheck} computes asteroid brightness using the $H$-$G$ magnitude system, which can be off by half magnitude at high phase angle \citep[i.e. low solar elongation;][]{Bernardi2009}. However, this is comparable in magnitude to other sources of uncertainty in our analysis (such as the change of limiting magnitude within a session), and thus will not affect our results in a significant way.

We then fit the brightness of known asteroids to the function, and derive $L=19.52\pm0.02$ (in $V$-band). This represents the point where the detection efficiency is $50\%$ of the maximum value, and refers to an average limiting magnitude throughout the survey. Although twilight fields are generally more susceptible to changing sky backgrounds, our airmass limit ($<2.92$) effectively ensures that the variation of sky background is within reasonable level in cloudless conditions. As a result, most of our images have limiting magnitudes of 19--20. The maximum efficiency is 70--80\% at the brighter end of the magnitude bin, not too far from the imaging filling factor of 87\%, attesting a decent performance of the survey throughout the period of performance.

\begin{figure}
\includegraphics[width=0.5\textwidth]{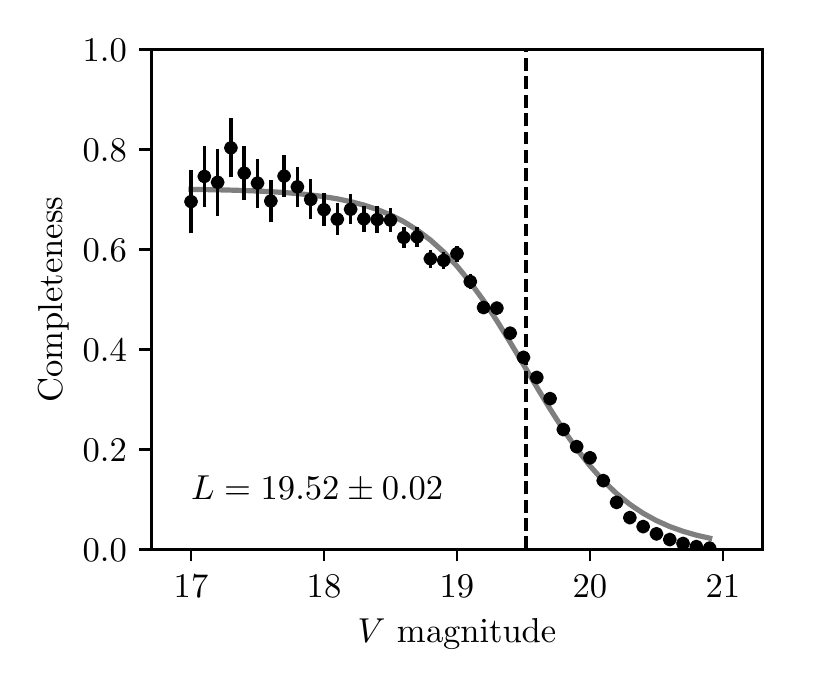}
\caption{Detection completeness of ZTF-TS as a function of $V$-band magnitude. The dashed line represents the limiting magnitude $L$, the point where the detection efficiency is $50\%$ of the maximum value. The solid line is the function defined by \citet{Denneau2013}.}
\label{fig:lm}
\end{figure}

We detected a total of 6 Atiras over the course of the survey, including 2 new Atiras (2019 AQ$_3$ and 2019 LF$_6$), all detected by {\tt ZMODE} (Table~\ref{tbl:obj}). This accounts for about a third of the known Atiras to date. (We also detected a pseudo-Atira, 2019 LL$_5$.) No Venus/Earth co-orbital asteroids are detected.

\begin{table*}
\begin{center}
\caption{Atiras and pseudo-Atiras detected by ZTF-TS. Objects discovered by the survey are in bold. No Venus/Earth co-orbital asteroids are detected. \label{tbl:obj}}
\begin{tabular}{cccccccc}
\hline
Designation & Date(s) of observation (UT) & Orbital type & $a$ (au) & $e$ & $i$ & $H$ & MPEC announcement \\
\hline
(163693) Atira & 2018 Dec. 5 & Atira & 0.74 & 0.32 & $25.6^\circ$ & 16.3 & - \\
(418265) 2008 EA$_{32}$ & 2019 Apr. 11 -- May 2 & Atira & 0.62 & 0.31 & $28.3^\circ$ & 16.4 & - \\
(434326) 2004 JG$_6$ & 2019 May 14 & Atira & 0.64 & 0.53 & $18.9^\circ$ & 18.4 & - \\
2018 JB$_3$ & 2018 May 29 -- Jun. 10 & Atira & 0.68 & 0.29 & $40.4^\circ$ & 17.6 & \citet{Ye2019b} \\
{\bf 2019 AQ$_3$ } & 2018 Dec. 22 -- 2019 Jan. 8 & Atira & 0.59 & 0.31 & $47.2^\circ$ & 17.6 & \citet{Buzzi2019} \\
{\bf 2019 LL$_5$ } & 2019 Jun. 5 -- 8 & Pseudo-Atira & 0.66 & 0.50 & $29.9^\circ$ & 18.2 & \citet{Ye2019c} \\
{\bf 2019 LF$_6$ } & 2019 Jun. 10 -- 19 & Atira & 0.56 & 0.43 & $29.5^\circ$ & 17.2 & \citet{Ye2019d} \\
\hline
\end{tabular}
\end{center}
\end{table*}

Apart from the confirmed discoveries in Table~\ref{tbl:obj}, we also discovered one new object, with a ZTF-assigned designation ZTF031C, that we were unable to confirm. Calculation by the JPL Scout system \citep{Farnocchia2015, Farnocchia2016} shows that ZTF031C has a $50\%$ probability of being an Atira-group asteroid.

\section{Discussion}

To understand the results, we employ the Near-Earth Object Population Observation Program \citep[NEOPOP;][]{Granvik2018a} to simulate the population of the Atiras and co-orbital asteroids. We generate 100 synthetic populations of Atiras and co-orbital asteroids using the NEOPOP Population Generator version 1.2.0, with absolute magnitudes of $10<H<20$. The range of $H$ is chosen by considering that (1) the largest NEO, (1036) Ganymed, has $H=9.5$; and (2) with a limiting magnitude of 19.5 (c.f. Figure~\ref{fig:lm}), ZTF-TS is unlikely to find many $H>20$ NEOs in the twilight direction (which experience brightness reduction due to the phase angle effect). We also assume a uniform magnitude slope $G=0.15$ for all synthetic asteroids following the statistical average derived by \citet{Bowell1989}. We then compare the ephemerides of each synthetic asteroid against the exposure catalog of ZTF-TS. The detection probability is computed using the sensitivity function derived in \S~\ref{sec:results}. The total number of synthetic asteroids detectable by ZTF-TS, $E(N)$, is therefore the sum of the probabilities that each synthetic asteroid $i$ has been detected at least once throughout the survey (with each survey session denoted as $j$):

\begin{equation}
E(N) = \sum_i{ \left[ 1-\prod_j{ \left(1-\epsilon_{i,j}\right) } \right]}
\end{equation}

It is worth noting about an implicit assumption here: synthetic asteroids that are detected at least once are considered as ``discoverable''. In reality, not every detection results in a confirmed discovery, with ZTF031C as a prominent example. We leave this out of our analysis, as it involves the follow-up efficiency of the global NEO network which is difficult to model.

The computed $E(N)$ for four populations of interest -- Atiras, Vatiras, Earth co-orbitals, and Venus co-orbitals -- along with the numbers of actual detections -- are tabulated in Table~\ref{tbl:mdl}, with uncertainties derived from the standard deviation from the model. We make the following remarks for each population of interest:

\begin{table*}
\begin{center}
\caption{The size of the population predicted by the NEOPOP model, as well as modeled and actual numbers of asteroids detectable by ZTF-TS. \label{tbl:mdl}}
\begin{tabular}{cccc}
\hline
Population & Total $N$ within $10<H<20$ & Modeled $E(N)$ & Observed $N$ \\
\hline
Atiras & $113\pm6$ & $13\pm2$ & 6\tablenotemark{a} \\
Vatiras & $18\pm1$ & $1.2\pm0.6$ & $0$ \\
Earth co-orbitals & $0.85\pm0.05$ & $0.04_{-0.04}^{+0.15}$ & $0$ \\
Venus co-orbitals & $0.46\pm0.03$ & $0.08_{-0.08}^{+0.22}$ & $0$ \\
\hline
\end{tabular}
\end{center}
\tablenotetext{a}{$N=8$ if we consider pseudo-Atira 2019 LL$_5$ and the unconfirmed object ZTF031C.}
\end{table*}

\paragraph{Atiras.} The model predicts $113\pm6$ Atiras with $H<20$, grossly in line with the number derived by \citet[][$\sim200\pm100$]{Zavodny2008}. The lower expectation value and smaller error bar is likely due to the inclusion of the disruptional removal of the near-Sun asteroids as well as $\sim10$~years worth of new survey data. $13\pm2$ Atiras in the survey data, while 6--8 Atiras are actually detected. The exact number depends on whether we consider pseudo-Atira 2019 LL$_5$ and the unconfirmed object ZTF031C as Atiras. (If we strictly follow the definition, pseudo-Atiras would not be Atiras.) Assuming Poison statistics, the probability of finding 8 or less Atiras is $\sim3\%$ ($0.4\%$ for finding 6 or less Atiras), but this is based on a small sample and may not have significant meaning. Interestingly, \citet{FuenteMarcos2019} suggests that 2019 AQ$_3$, a large Atira found in the course of the survey, is an outlier in the context of the same NEOPOP model. These seemingly contradicting conclusions suggest that our interpretation is limited by small statistics.

\paragraph{Vatiras.} Vatiras, a portmanteau of {\it Venus} and {\it Atiras}, are a hypothetical asteroid group first proposed by \citet{Greenstreet2012}. Similar to Atiras to the Earth, Vatiras are asteroids whose orbits are entirely confined by the orbit of Venus. NEOPOP predicts $1.2\pm0.6$ Vatiras in the ZTF-TS survey data. Even though none has been actually detected, the model expectation suggests that the first Vatira may soon be found. In fact, 2019 AQ$_3$ has the smallest aphelion distance of any known asteroids and is just outside the realm of Vatiras ($Q=0.774$~au for 2019 AQ$_3$ versus Vatira definition of $Q<0.718$~au). Dynamical study shows that 2019 AQ$_3$ may periodically migrate between Vatira and Atira orbits due to the close encounters to Venus \citep{FuenteMarcos2019}.

\paragraph{Earth and Venus Co-orbital asteroids.} It is perhaps unsurprising that the survey detected zero co-orbital asteroids, given that large ($H<20$) co-orbital asteroids appear to be either very rare or non-existent. The non-detection is consistent with the near-zero expectation predicted by the model. We note that the expectation of detectable Earth co-orbitals is lower than that of Venus co-orbitals even though the population of Earth co-orbitals is actually larger than the Venus co-orbitals \citep{Morais2006}. This is because ZTF-TS targets fields that are closer to the Sun than where most Earth co-orbitals can be found ($\sim60^\circ$ from the Sun).

\begin{figure}
\includegraphics[width=0.5\textwidth]{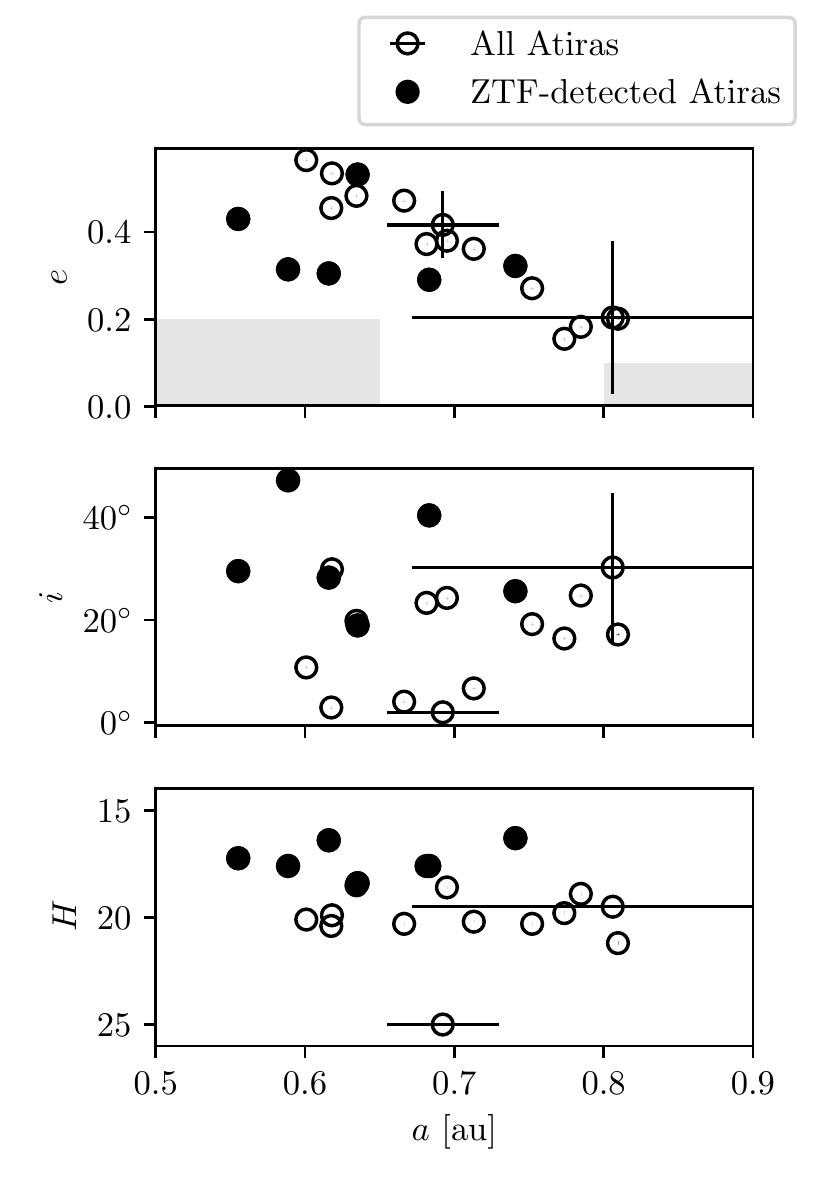}
\caption{Orbital elements ($a$, $e$, $i$) and physical parameters ($H$) of known Atiras. Objects detected by ZTF are plotted in filled circles while others are plotted in open circles. Uncertainties are also plotted, but are too small for most objects to be noticable. Shaded areas are the ``islands of dynamical stability'' identified by \citet{Ribeiro2016}.}
\label{fig:atiras}
\end{figure}

Figure~\ref{fig:atiras} shows the orbital distribution of the known Atiras, with those detected by ZTF highlighted. Also shown in the figures are the ``islands of stability'' identified by \citep{Ribeiro2016} where $e<0.2$. We see that ZTF-TS is sensitive to large ($H<18$) objects on moderately eccentric ($e>0.3$) and inclined ($i>20^\circ$) orbits, likely due to its survey strategy and geographic location: by going as close to the Sun as possible, ZTF-TS is insensitive to small objects that come close to the Earth, and the mid-latitude geographic location of the survey reduces its sensitivity to low-inclination objects that stay near the ecliptic plane. The ``islands of stability'', a set of hypothesized regions between the orbits of Mercury, Venus and Earth, contain objects with low eccentricities ($e<0.2$) which make them difficult to detect. However, those that are on moderately-inclined orbits are still somewhat easier to detect for mid-latitude surveys such as ZTF-TS. Atiras 2019 AQ$_3$ ($i=47^\circ$) and 2019 LF$_6$ ($i=30^\circ$), for example, are likely on the edge of the island of stability between Mercury and Venus. Recent dynamical investigations have shown their longevity in this region \citep{FuenteMarcos2018, FuenteMarcos2019, FuenteMarcos2019a}.

\section{Conclusions}

We have presented a twilight survey in search of NEOs on or within the Earth's orbit, carried out on the recently commissioned ZTF project during its first year of operation. ZTF-TS operates on skies with solar elongations of $35^\circ$--$60^\circ$. Over a total of 82 nights (40 evening sessions and 62 morning sessions), we detected six unambiguous Atiras, including two previously unknown ones. No Vatiras or Earth/Venus co-orbital asteroids are detected. We plan to continue the survey through the end of the ZTF survey, which is expected to conclude at the end of 2020.

Due to its mid-latitude geographic location, ZTF-TS favors mid-ecliptic latitude skies. This helps the detection/discovery of objects on moderately-inclined orbits, such as 2019 AQ$_3$ and 2019 LF$_6$, though it also introduces a bias in the sample. We expect more objects to be found in this largely uncharted region, possibly including the hypothetical Vatiras. However, NEO population model predicts $N=113\pm6$ for the $H<20$ Atira population, which means that a survey like ZTF-TS only detects $\sim5\%$ of the total population in a 7-month survey. It will take a decade or more for a survey like ZTF-TS to detect most of the $H<20$ Atiras, assuming the same observing strategy and conditions.

The non-detection of Venus and Earth co-orbital asteroids is consistent with previous predictions. This highlights the challenge of searching for these asteroids. Admittedly, ZTF-TS operates in skies that are too close to the Sun to detect the Earth co-orbitals, but the ``sweet-spots'' for Earth co-orbitals are already regularly patrolled by other NEO telescopes, including the ones that are more sensitive than ZTF (e.g. PANSTARRS). Only 13 Earth co-orbital asteroids have been found so far, with the largest member being (138852) 2000 WN$_{10}$ ($H=20.2$). No Venus co-orbital asteroid has ever been found. These objects likely exist in large numbers in smaller diameters, but detection of such objects requires deeper surveys.

Future surveys to search for NEOs interior to the Earth's orbit, such as the twilight survey proposed for the Large Synoptic Survey Telescope \citep{Seaman2018} and NEOCam's space-based survey from the first Earth-Sun Lagrange point \citep{Mainzer2018}, will likely boost the discovery of Atiras and co-orbital asteroids. We do however note that these surveys will operate at solar elongations $>45^\circ$, which will somewhat limit their effectiveness in detecting Vatiras (which only reach a maximum solar elongation of $47^\circ$) and other low-eccentricity objects. Nevertheless, these surveys will provide a critical stepping stone towards the understanding of these populations.




\acknowledgments

We thank an anonymous referee for comments. Q.-Z. Ye acknowledges support by the GROWTH (Global Relay of Observatories Watching Transients Happen) project funded by the National Science Foundation PIRE (Partnership in International Research and Education) program under Grant No. 1545949. We also thank Dave Tholen for discussion in the early phase of ZTF-TS and Bryce Bolin for comments. The work of W.-H. Ip was supported in part by grant No. 107-2119-M-008-012 of MOST, Taiwan, and grant No. 119/2017/A3 of FCDT of Macau, MSAR. D. Farnocchia conducted this research at the Jet Propulsion Laboratory, California Institute of Technology, under a contract with NASA.

Based on observations obtained with the Samuel Oschin Telescope 48-inch Telescope at the Palomar Observatory as part of the Zwicky Transient Facility project. ZTF is supported by the National Science Foundation under Grant No. AST-1440341 and a collaboration including Caltech, IPAC, the Weizmann Institute for Science, the Oskar Klein Center at Stockholm University, the University of Maryland, the University of Washington, Deutsches Elektronen-Synchrotron and Humboldt University, Los Alamos National Laboratories, the TANGO Consortium of Taiwan, the University of Wisconsin at Milwaukee, and Lawrence Berkeley National Laboratories. Operations are conducted by COO, IPAC, and UW.

\facilities{PO:1.2m}
\software{astcheck, Matplotlib \citep{Hunter2007}, NEOPOP \citep{Granvik2018a}, sbpy \citep{Mommert2018}}

\end{CJK*}
\bibliographystyle{aasjournal}
\bibliography{ms}



\end{document}